\documentclass[twocolumn,tbtags,superscriptaddress,floatfix,showpacs]{revtex4-1}
\usepackage{amssymb}
\usepackage{graphicx,amsmath,subcaption}
\usepackage{bm}
\usepackage{times}
\usepackage{latexsym}
\usepackage{breqn}
\usepackage{etoolbox}
\usepackage{textcomp}
\usepackage{mwe}

\begin{document}
\title{Universal tool for single-photon circuits: quantum router design }

\begin{abstract}
We demonstrate that the non-Hermitian Hamiltonian approach can be used as a universal tool to design and
describe a performance of single photon quantum electrodynamical circuits (cQED). As an example of the validity
of this method, we calculate a novel six port quantum router, constructed from 4 qubits and 3 open-waveguides.
We have got analytical expressions, which describe the transmission and reflection coefficients of a single photon
in general form taking into account the non-uniform qubit’s parameters. We show that, due to naturally derived
interferences, it is possible to tune the probability of photon detection in different ports in-situ.
\end{abstract}


\makeatletter
\preto\maketitle{%
  \begingroup\lccode`~=`,
  \lowercase{\endgroup
  \let\saved@breqn@active@comma~
  \let~}\active@comma 
}
\appto\maketitle{
  \begingroup\lccode`~=`,
  \lowercase{\endgroup
  \let~}\saved@breqn@active@comma 
}
\makeatother

\author{Sultanov A. N.}
\affiliation{Novosibirsk State Technical University, Novosibirsk, Russia}
\author{Greenberg Ya. S.}
\affiliation{Novosibirsk State Technical University, Novosibirsk, Russia}
\author{Mutsenik E. A.}
\affiliation{Novosibirsk State Technical University, Novosibirsk, Russia}
\author{Pitsun D. K.}
\affiliation{Novosibirsk State Technical University, Novosibirsk, Russia}
\date{\today}
\author{Il'ichev E.}
\affiliation{Leibniz Institute of photonic technology, Jena, Germany}
\affiliation{Russian Quantum Center, Skolkovo, Russia}
\maketitle

\section*{Introduction}
	Quantum circuits represent an important part of a rapidly developing research area, which includes quantum information transfer and processing. In general, these circuits could be presented as a set of quantum nodes, interconnected by quantum channels.
Photons propagation in these circuits is accompanied by the very low coherence losses even for long distances, which makes them the prime contender for the quantum information carriers. It is well known, that for some applications of quantum cryptography and
data transfer it is necessary to have a single photon \cite{1}. It is obvious that for the fully quantum device we should investigate some instrument, which allows controlling the interactions between different quantum nodes.
In the past decade, a big amount of such devices has been offered, and they were named as quantum routers or switches. So, generally, the task of design a quantum router for practical application is the problem of single photon scattering.
Naturally, most of these devices were realized for the optical range \cite{2, 3, 4, 5, 6, 7, 8, 9, 10, 11, 12}, and only several works were addressed to the ones for microwave range \cite{13,14,15,16,17}. Basically, these devices could be divided into two main groups: exploiting multilevel atoms \cite{4,6,8,16} and employing two-level quantum system (TLS) \cite{9,11,12,13,14,15,17}.

It is worth noting that these devices were also investigated for 3D optic lattice \cite{10}, utilizing nanomechanical systems \cite{2} and as a combination of standard quantum gates \cite{18}. Theoretically, most of the structures were described by making use of standard methods:
direct solving of Schrodinger equations or Heisenberg representation approach. In the first case, the problem of the systems with three (or less) elements (qubits+cavities+waveguides) is solved by means of discretization of waveguides (constructed from coupled resonance cavities) \cite{4,15,16}. For this case the Laplace transformations \cite{19} or the real-space approach \cite{8,9} are used. In the Heisenberg representation approach, the input-output formalism is utilized for the similar systems with a modest complexity \cite{3,6,7,11,12,13}. It worth to mention that results of numerically solved master-equation for the device constructed of 2 scatterers, waveguide and two-mode cavity (considered as 5 element system), is presented in \cite{3}. Up to date (to our knowledge) this is the most complex investigated system.
    In this paper, we demonstrate that non-Hermitian Hamiltonian approach can be used as a universal tool for describing and designing single-photon quantum electrodynamic (QED) circuits. This method was firstly developed in the nuclear physics \cite{20,21}, and was later adapted to QED schemes \cite{22}. It presents a universal method to a full quantum mechanical description of the scattering and statistical parameters of any QED chain. It was successfully implemented to describe photon transport through the different multi qubit structures \cite{22,23,24,25,26,27}. This method has a set of advantages, and the most important is the natural account for the retardation effect and the derivation of the explicit form for the whole system’s wave functions. The method allows accounting for non-uniform parameters, which is very important for real solid-state devices, which based on artificial elements. Generally, this method is appropriate to arbitrary quantum circuits, consisting of transmission lines, cavities and TLS, if we are interested in single photon scattering probabilities. Also, it can be used as a universal instrument for analyzing and designing the circuits in a majority of QED structures. Of course, further study is required to generalize the method to multilevel atoms and many photon scattering problems in quantum circuits. In comparison with other works, where the interference was introduced as an external condition (for example, standing wave condition coming from classical optics), here the interference appears naturally from the standard light-matter interaction Hamiltonian. Moreover, it is not so obvious how to implement the conventional methods from quantum optics for real QED circuits.

    The best demonstration of the method’s power is to apply it for multichannel QED circuit. We chose a quantum router because it is generally a system with a discrete spectrum (it could be constructed from two-level or multilevel atoms, oscillators and so on), which is connected with a continuum by several decay channels (open waveguides, for example). Using this method we present a novel variant of the broadband scalable quantum router, which can be realized both in microwave range (utilizing superconducting coplanar waveguides and solid-state qubits), and in an optical range (utilizing quantum dots). Additionally note, that proposed and calculated device has the largest number of ports in comparison with known ones.
The main requirements for quantum routers of single photons are presented at \cite{18}: 1) both the signal and control information have to be stored in quantum objects; 2) the signal has to be unchanged under the routing operation; 3) the router has to be able to route the signal into a coherent superposition; 4) the router has to work without any need for post-selection; 5) in order to optimize the resources of the quantum network, every individual qubit has to control a single photon signal. In the context of a usable device, we want to add the requirement of a broadband access, which is impossible to realize for the system where resonators and cavities are being used \cite{2,3,4,7,8,11,12,13,16}. Another important requirement is a scalability.
The scheme of the proposed device is shown in Fig 1. It presents a 6-port device, containing 4 two-level atoms and three open-waveguides. One waveguide is connected with two others. The similar scheme was proposed in theoretical work [19], where the interaction between open waveguides is controlled by a two-level atom. However, the main advantages of our device are the accounting for the distances between TLS, clear scalability, near unity routing quality, the opportunity to create a coherent superposition state and absence of necessity to use additional circulators, which are presented in all works with open waveguides. So, the last advantage automatically fulfills the condition 4) from \cite{18}. Also, non-use of any cavities gives us the advantage of broadbandness, and by help of the non-Hermitian approach, we show, that the routing arises naturally on quantum mechanical fundamentals. Moreover, we show that our method accounts for the interaction in all orders of coupling strength between TLS and scattering photon. This scheme has two different operating modes: (i) a six-port non-symmetric router and (ii) a 4-port symmetric device.

The paper is arranged as follows. In Sec. I we present the circuit scheme followed by basics of its theoretical description in terms of non-Hermitian Hamiltonian approach. In Sec. II we focus on the analytical equations for routing probabilities and also present the exact expressions of system’s stationary wave function. There we show how to get the scattering coefficients from these wave functions and analyze some limit cases of the scattering parameters. In Sec. III we introduce one-dimensionality conditions and implement them to the scheme. In Sec IV. simulation results are presented, and two operating modes are shown. Here we demonstrate how to control the probabilities through TLS tuning. Finally, in Sec. V we summarize our results and discuss the research directions which could be done on the base of our results.

\section{The system and method description}
The system under investigation presents three waveguides and four TLS, the last one we will name below as a qubit. Each waveguide contains two qubits, and the system is constructed in such way that three waveguides are interconnected through two qubits in $B$ waveguide (see Fig 1a). There two qubits Q3 and Q4  are coupled only to waveguide $A$ and waveguide $C$ correspondingly, and qubits Q1 and Q2 are coupled to waveguide $B$ and provide coupling to waveguides $C$ and $A$, correspondingly.
\begin{figure}[h]
  \begin{subfigure}[b]{0.4\textwidth}
    \includegraphics[width=\textwidth]{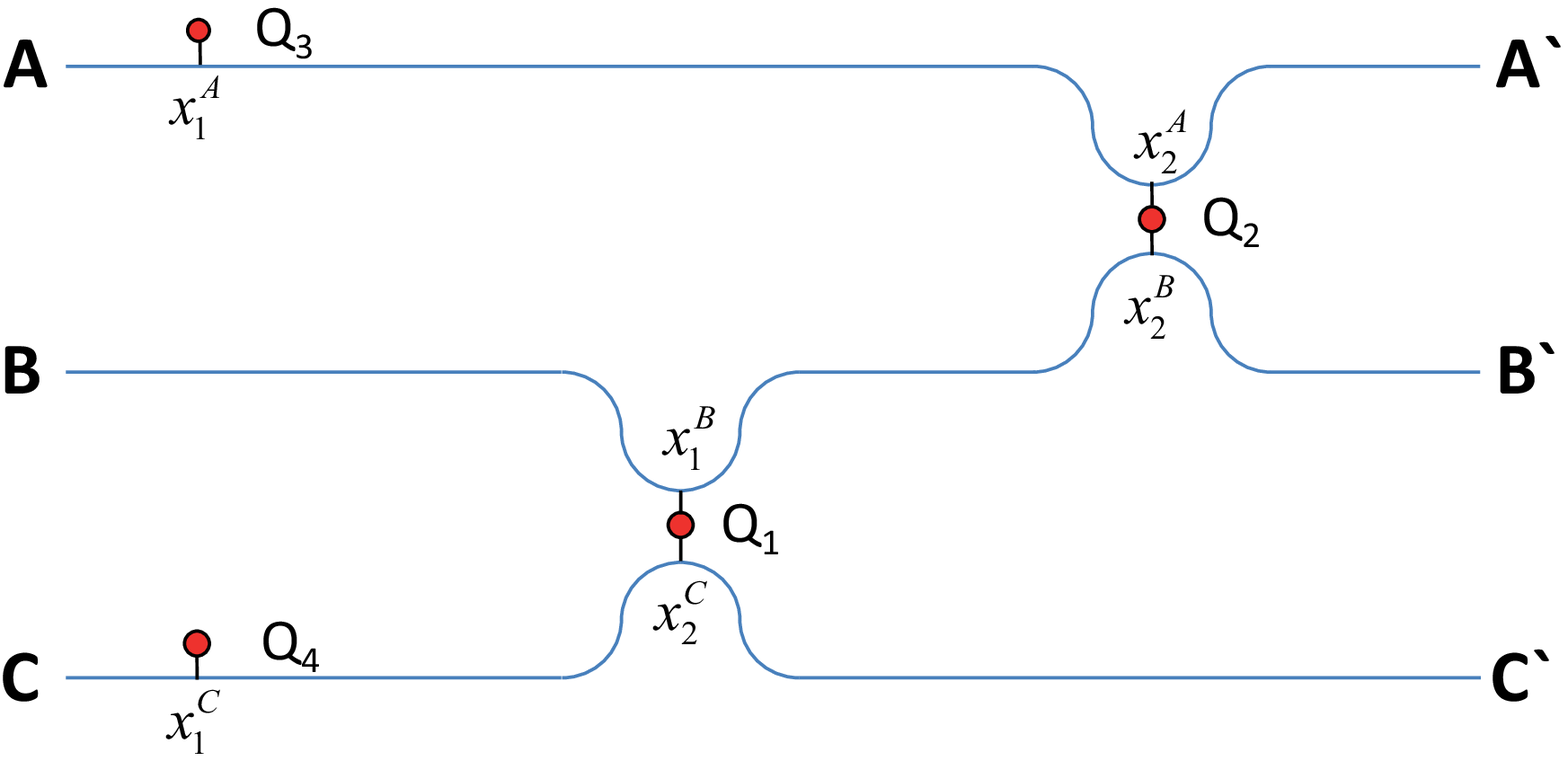}
    \caption{}
    \label{fig1a}
  \end{subfigure}
  \hfill 
  \begin{subfigure}[b]{0.4\textwidth}
    \includegraphics[width=\textwidth]{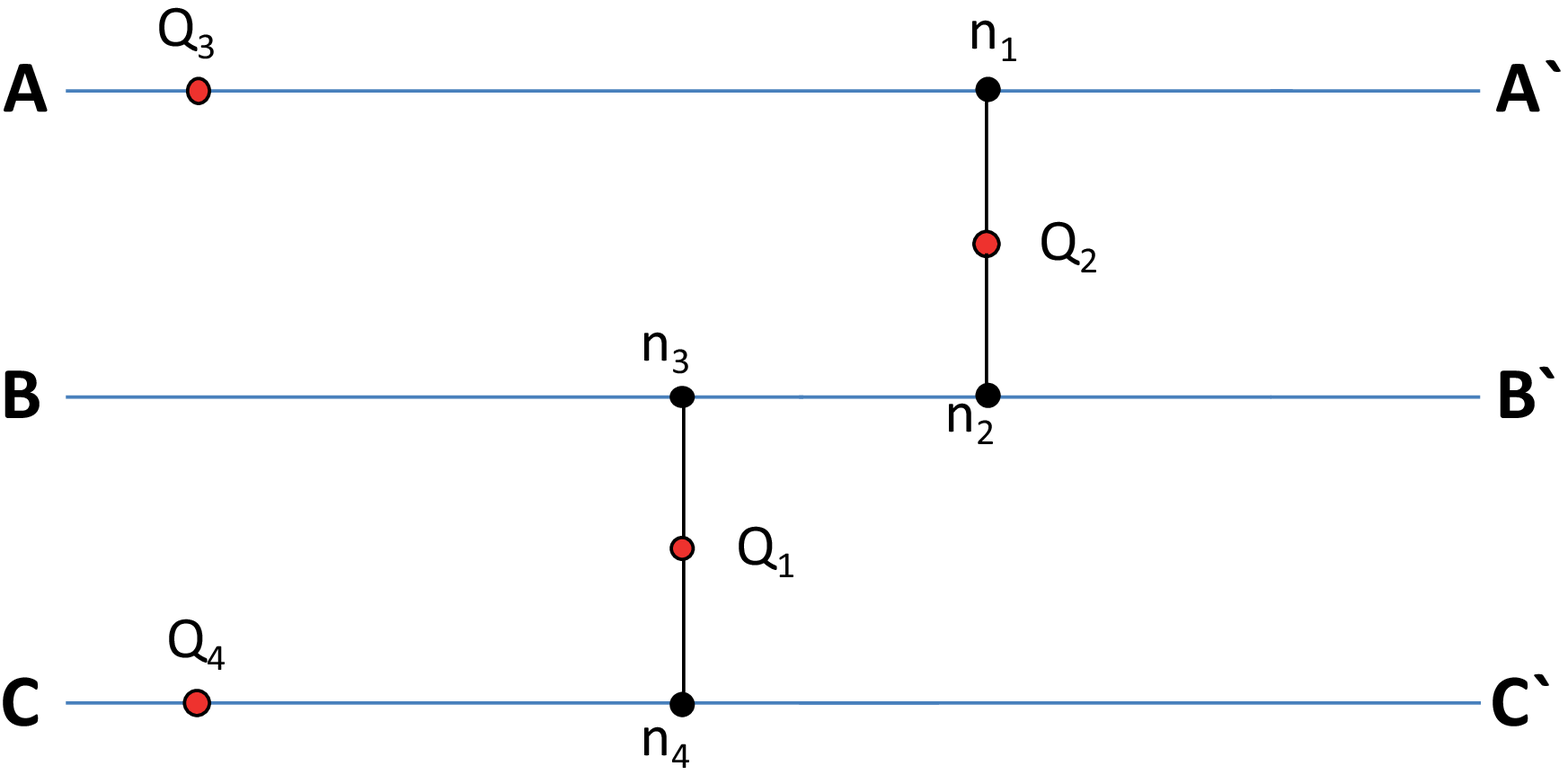}
    \caption{}
    \label{fig1b}
  \end{subfigure}
\caption{The scheme of quantum router: a) sketch of the system constructed from 3 waveguides ($A, B, C$) and 4 qubits at coordinates $x_I^J$  , where $I$ means position at the $J^{th}$ waveguide ($J = A, B, C$). b) Simplified diagram of the system}
\label{fig1}
\end{figure}
The full Hamiltonian of the system is written as follows:
\begin{dmath}\label{full}
H = \sum\limits_{k,l = a,b,c} {\hbar \omega _k^l\hat l_k^\dag \hat l_k^{}}  + \sum\limits_{i = 1}^4 {\frac{1}{2}\hbar {\Omega _i}\left( {1 + \hat \sigma _z^{\left( i \right)}} \right)}  + \sum\limits_k^{} {\hbar \xi _A^3\left( {\hat a_k^\dag \hat \sigma _ - ^{\left( 3 \right)}{e^{ - jkx_1^A}} + {{\hat a}_k}\hat \sigma _ + ^{\left( 3 \right)}{e^{jkx_1^A}}} \right)} 
 + \sum\limits_k^{} {\hbar \xi _A^2\left( {\hat a_k^\dag \hat \sigma _ - ^{\left( 2 \right)}{e^{ - jkx_2^A}} + {{\hat a}_k}\hat \sigma _ + ^{\left( 2 \right)}{e^{jkx_2^A}}} \right)}  + \sum\limits_k^{} {\hbar \xi _B^1\left( {\hat b_k^\dag \hat \sigma _ - ^{\left( 1 \right)}{e^{ - jkx_1^B}} + {{\hat b}_k}\hat \sigma _ + ^{\left( 1 \right)}{e^{jkx_1^B}}} \right)} 
 + \sum\limits_k^{} {\hbar \xi _B^2\left( {\hat b_k^\dag \hat \sigma _ - ^{\left( 2 \right)}{e^{ - jkx_2^B}} + {{\hat b}_k}\hat \sigma _ + ^{\left( 2 \right)}{e^{jkx_2^B}}} \right)}  + \sum\limits_k^{} {\hbar \xi _C^1\left( {\hat c_k^\dag \hat \sigma _ - ^{\left( 4 \right)}{e^{ - jkx_1^C}} + {{\hat c}_k}\hat \sigma _ + ^{\left( 4 \right)}{e^{jkx_1^C}}} \right)}  + \sum\limits_k^{} {\hbar \xi _C^2\left( {\hat c_k^\dag \hat \sigma _ - ^{\left( 1 \right)}{e^{ - jkx_2^C}} + {{\hat c}_k}\hat \sigma _ + ^{\left( 1 \right)}{e^{jkx_2^C}}} \right)} 
\end{dmath}
where $\hat l_k^\dag \left( {\hat l_k^{}} \right)$  are the bosonic operators of creation (annihilation) photons with wave vector $k$ (and with energy $\hbar \omega _k^l$  ) in $l^{th}$ waveguide; $\hat \sigma _z^{\left( i \right)} = \left| {{e_i}} \right\rangle \left\langle {{e_i}} \right| - \left| {{g_i}} \right\rangle \left\langle {{g_i}} \right|$ - Pauli spin operator, where $\left| {{e_i}} \right\rangle $ ($\left| {{g_i}} \right\rangle $ ) are the excited (ground) state of $i^{th}$ qubit; the interaction between waveguide photons and qubits are described 
through Jaynes-Cummings model, where  $\hat \sigma _ - ^{\left( i \right)} = \left| {{g_i}} \right\rangle \left\langle {{e_i}} \right|$ ($\hat \sigma _ + ^{\left( i \right)} = \left| {{e_i}} \right\rangle \left\langle {{g_i}} \right|$ ) are lowering (raising) operators for $i^{th}$ qubit and $\xi _J^i$  is the coupling between $J^{th}$ waveguide and $i^{th}$ qubit;  $\hbar $ is the Planck constant, and hereafter we take   $\hbar  = 1$.
For one photon routing we restrict our states basis to the one-excitation states, so we introduce them as follows:
\begin{equation}\label{ext_st}
    \begin{aligned}
\left| A \right\rangle  = \left| {{k_A}{{,0}_B}{{,0}_C}} \right\rangle  \otimes \left| G \right\rangle \\
\left| B \right\rangle  = \left| {{0_A},{k_B}{{,0}_C}} \right\rangle  \otimes \left| G \right\rangle \\
\left| C \right\rangle  = \left| {{0_A}{{,0}_B},{k_C}} \right\rangle  \otimes \left| G \right\rangle     
    \end{aligned}
\end{equation}
where the $\left| G \right\rangle  = \left| {{g_1},{g_2},{g_3},{g_4}} \right\rangle $   is the ground state of all qubits, so $\left| J \right\rangle $  ($J=A,B,C$) describes the situation where we have one photon in $J^{th}$ waveguide and all qubits in the ground states. Then this single photon can be absorbed by one of four qubits, leaving the waveguides empty, so we define these states as:
\begin{equation}\label{int_st}
\begin{aligned}
\left| 1 \right\rangle  = \left| {{e_1},{g_2},{g_3},{g_4}} \right\rangle  \otimes \left| {{0_{vac}}} \right\rangle \\
\left| 2 \right\rangle  = \left| {{g_1},{e_2},{g_3},{g_4}} \right\rangle  \otimes \left| {{0_{vac}}} \right\rangle \\
\left| 3 \right\rangle  = \left| {{g_1},{g_2},{e_3},{g_4}} \right\rangle  \otimes \left| {{0_{vac}}} \right\rangle \\
\left| 4 \right\rangle  = \left| {{g_1},{g_2},{g_3},{e_4}} \right\rangle  \otimes \left| {{0_{vac}}} \right\rangle 
\end{aligned}
\end{equation}
where $\left| {{0_{vac}}} \right\rangle $  is the photonic vacuum state (such that $l_k^\dag \left| {{0_{vac}}} \right\rangle  = \left| {{k_J}} \right\rangle $  respectively for each waveguide), and  $\left| i \right\rangle $ (where $i=1..4$) describes the state when we have excited $i^{th}$ qubit.
It is obvious, that the states (\ref{ext_st}) have continuous energy spectra, due to arbitrary photon’s frequency in an open waveguide, and the states (\ref{int_st}) have discrete energy spectrum. It allows us to define two different groups of  Hilbert space’ states with continuous and discrete spectrum. This subdivision is proper for easy calculation in a frame of the non-Hermitian Hamiltonian approach (but generally the projection procedure can use any subdivision of space). This way, following the non-Hermitian Hamiltonian approach \cite{22}, we introduce the projection operators:
\begin{equation}\label{proj}
    \begin{array}{l}
\hat P = \frac{{{L_A}}}{{2\pi }}\int {d{k_A}\left| A \right\rangle \left\langle A \right|}  + \frac{{{L_B}}}{{2\pi }}\int {d{k_B}\left| B \right\rangle \left\langle B \right|} \\ + \frac{{{L_C}}}{{2\pi }}\int {d{k_C}\left| C \right\rangle \left\langle C \right|} \\
\hat Q = \left| 1 \right\rangle \left\langle 1 \right| + \left| 2 \right\rangle \left\langle 2 \right| + \left| 3 \right\rangle \left\langle 3 \right| + \left| 4 \right\rangle \left\langle 4 \right|
    \end{array}
\end{equation}
which obey the following equations $\hat P\hat Q = \hat Q\hat P = 0;\hat P\hat P = \hat P;\hat Q\hat Q = \hat Q;\hat P + \hat Q = 1$ . The last equations present the fullness of chosen basis for one-excitation states. We want to describe probabilities of transitions between states (\ref{ext_st}) through all trajectories including internal states (\ref{int_st}) by use of the non-Hermitian effective Hamiltonian. This Hamiltonian is defined fully in a basis of internal states (\ref{int_st}), describing decay of these states due to coupling to the continuum. Here we leave out routine calculations, which are similar to ones from \cite{22}, and write the effective Hamiltonian and full system’s wave functions as follows:

\begin{equation}
{\hat H_{eff}} = {\hat H_{QQ}} + {\hat H_{QP}}\frac{1}{{E - {{\hat H}_{PP}} + i\varepsilon }}{\hat H_{PQ}};
\end{equation}

\begin{dmath}
\left| {{\Psi _{in}}} \right\rangle  = \left| {in} \right\rangle  + \frac{1}{{E - {{\hat H}_{eff}}}}{\hat H_{QP}}\left| {in} \right\rangle  + \frac{1}{{E - {{\hat H}_{PP}}}}{\hat H_{PQ}}\frac{1}{{E - {{\hat H}_{eff}}}}{\hat H_{QP}}\left| {in} \right\rangle 
\end{dmath}
where ${\hat H_{XY}} = \hat X\hat H\hat Y$  ( $\hat X,\hat Y = \hat P,\hat Q$ ) is projection of the full Hamiltonian (\ref{full}) and  $\left| {in} \right\rangle $ presents the system’s state before the photon scattered at multiqubit internal system, which can be expressed through the states (\ref{int_st}) with defined initial wave vector ${k_0} = \frac{\omega }{{{\nu _g}}}$ (hereafter $\omega$ is the scattering photon's frequency) and initial state’s energy $E$, i.e. $\left| {in} \right\rangle  = \left| {{A_0}} \right\rangle ,\left| {{B_0}} \right\rangle ,\left| {{C_0}} \right\rangle $ . Since the chosen basis of states is full, we use the fullness property:
\begin{dmath*}
\left| {{\Psi _{in}}} \right\rangle  = \left| {in} \right\rangle  + \frac{1}{{E - {{\hat H}_{eff}}}}\left( {\hat P + \hat Q} \right){\hat H_{QP}}\left| {in} \right\rangle  + \frac{1}{{E - {{\hat H}_{PP}}}}\left( {\hat P + \hat Q} \right){\hat H_{PQ}}\left( {\hat P + \hat Q} \right)\frac{1}{{E - {{\hat H}_{eff}}}}\\
\left( {\hat P + \hat Q} \right){\hat H_{QP}}\left| {in} \right\rangle 
\end{dmath*}
 where the second term describes only internal system behavior, which is in undetectable in our case (we detect only the photon), and this substitution with Eq. (\ref{proj}) leads us to following form of full system's wave function:
 \begin{dmath}\label{gen_fun}
\left| {{\Psi _{in}}} \right\rangle  = \left| {in} \right\rangle  + \sum\limits_{n,m = 1}^4 {\left| n \right\rangle } {R_{nm}}\left\langle m \right|{\hat H_{QP}}\left| {in} \right\rangle  + \int {dq\sum\limits_{J = A,B,C} {\sum\limits_{n,m = 1}^4 {\frac{{\left| {{q_J}} \right\rangle }}{{E - {E_J}\left( q \right)}}\left\langle J \right|{{\hat H}_{PQ}}\left| n \right\rangle \\
 {R_{nm}}\left\langle m \right|{{\hat H}_{QP}}\left| {in} \right\rangle } } } 
\end{dmath}
where ${R_{mn}} = \left\langle m \right|\frac{1}{{E - {H_{eff}}}}\left| n \right\rangle $ . 

From (\ref{gen_fun}) it is clear that for each initial state $\left| {in} \right\rangle $ there are some probabilities to be transformed to one of 3 different states (summation over $J$). Probabilitiy of each scenario is defined by the different combinations of interaction through the internal states (summation over $n, m$). The strength of these interaction depends on couplings between the internal and initial states (term $\left\langle m \right|{\hat H_{QP}}\left| {in} \right\rangle$), between internal all external states in general (terms $\left\langle J \right|{\hat H_{PQ}}\left| n \right\rangle$) and  also depends on effective interaction between internal states (term ${R_{nm}}$)
The effective Hamiltonian of the system could be expressed in a matrix form in the basis (\ref{int_st}) as:
\begin{widetext}
\begin{equation}\label{h_eff}
{\hat H_{eff}} = \left( {\begin{array}{*{20}{c}}
{{\Omega _1} - j{\Gamma _{B1}} - j{\Gamma _{C1}}}&{ - j\sqrt {{\Gamma _{B1}}{\Gamma _{B2}}} {e^{jk\left| {x_1^B - x_2^B} \right|}}}&0&{ - j\sqrt {{\Gamma _{C1}}{\Gamma _{C4}}} {e^{jk\left| {x_1^C - x_2^C} \right|}}}\\
{ - j\sqrt {{\Gamma _{B1}}{\Gamma _{B2}}} {e^{jk\left| {x_1^B - x_2^B} \right|}}}&{{\Omega _2} - j{\Gamma _{B2}} - j{\Gamma _{A2}}}&{ - j\sqrt {{\Gamma _{A1}}{\Gamma _{A3}}} {e^{jk\left| {x_1^A - x_2^A} \right|}}}&0\\
0&{ - j\sqrt {{\Gamma _{A1}}{\Gamma _{A3}}} {e^{jk\left| {x_1^A - x_2^A} \right|}}}&{{\Omega _3} - j{\Gamma _{A3}}}&0\\
{ - j\sqrt {{\Gamma _{C1}}{\Gamma _{C4}}} {e^{jk\left| {x_1^C - x_2^C} \right|}}}&0&0&{{\Omega _4} - j{\Gamma _{C4}}}
\end{array}} \right)
\end{equation}
\end{widetext}
where  ${\Gamma _{Ji}}$($J=A,B,C$ and $i=1..4$) describes the $i^{th}$ qubit decay rate to the $J^{th}$ waveguide, and it can be expressed through the couplings ${\Gamma _{Ji}} = \frac{{{L_J}{{\left( {\xi _J^i} \right)}^2}}}{{v_g^J}}$ , where the ${L_J}$   and $v_g^J$  are the length of $J^{th}$ waveguide and photon wave’s group velocity in$J^{th}$ waveguide with a linear dispersion, consequently.

\section{Solutions and transmission probabilities}
Because the initial state could be prepared as one of three states (see Eq. (\ref{ext_st})), according to (7) we get three different wave functions. These wave functions, after some routine algebra, (omitting details), allow us to find nine solutions (three outcomes per each initial state) in configuration space:
        
\begin{widetext}
\begin{subequations}\label{psi_f}
\begin{dgroup*}
        \begin{dmath}
         \left\langle {{x_A}} \right.\left| {{\Psi _A}} \right\rangle  = {e^{j{k_0}x}} - j\sqrt {{\Gamma _{A2}}} {e^{j{k_0}x_2^A}}\left( {{R_{22}}\sqrt {{\Gamma _{A2}}} {e^{j{k_0}\left| {x - x_2^A} \right|}} + {R_{32}}\sqrt {{\Gamma _{A3}}} {e^{j{k_0}\left| {x - x_1^A} \right|}}} \right) - j\sqrt {{\Gamma _{A3}}} {e^{j{k_0}x_1^A}}\left( {{R_{23}}\sqrt {{\Gamma _{A2}}} {e^{j{k_0}\left| {x - x_2^A} \right|}} + {R_{33}}\sqrt {{\Gamma _{A3}}} {e^{j{k_0}\left| {x - x_1^A} \right|}}} \right);
\label{psi_aa}
        \end{dmath}
    \begin{dmath}
\left\langle {{x_B}} \right.\left| {{\Psi _A}} \right\rangle  =  - j\sqrt {{\Gamma _{A2}}} {e^{j{k_0}x_2^A}}\left( {{R_{12}}\sqrt {{\Gamma _{B1}}} {e^{j{k_0}\left| {x - x_1^B} \right|}} + {R_{22}}\sqrt {{\Gamma _{B2}}} {e^{j{k_0}\left| {x - x_2^B} \right|}}} \right) - j\sqrt {{\Gamma _{A3}}} {e^{j{k_0}x_1^A}}\left( {{R_{13}}\sqrt {{\Gamma _{B1}}} {e^{j{k_0}\left| {x - x_1^B} \right|}} + {R_{23}}\sqrt {{\Gamma _{B2}}} {e^{j{k_0}\left| {x - x_2^B} \right|}}} \right);
\label{psi_ab}
    \end{dmath}
\begin{dmath}
\left\langle {{x_C}} \right.\left| {{\Psi _A}} \right\rangle  =  - j\sqrt {{\Gamma _{A2}}} {e^{j{k_0}x_2^A}}\left( {{R_{12}}\sqrt {{\Gamma _{C1}}} {e^{j{k_0}\left| {x - x_2^C} \right|}} + {R_{42}}\sqrt {{\Gamma _{C4}}} {e^{j{k_0}\left| {x - x_1^C} \right|}}} \right) - j\sqrt {{\Gamma _{A3}}} {e^{j{k_0}x_1^A}}\left( {{R_{13}}\sqrt {{\Gamma _{C1}}} {e^{j{k_0}\left| {x - x_2^C} \right|}} + {R_{43}}\sqrt {{\Gamma _{C4}}} {e^{j{k_0}\left| {x - x_1^C} \right|}}} \right);
\label{psi_ac}
    \end{dmath}
\begin{dmath}
\left\langle {{x_A}} \right.\left| {{\Psi _B}} \right\rangle  =  - j\sqrt {{\Gamma _{B1}}} {e^{j{k_0}x_1^B}}\left( {{R_{21}}\sqrt {{\Gamma _{A2}}} {e^{j{k_0}\left| {x - x_2^A} \right|}} + {R_{31}}\sqrt {{\Gamma _{A3}}} {e^{j{k_0}\left| {x - x_1^A} \right|}}} \right) - j\sqrt {{\Gamma _{B2}}} {e^{j{k_0}x_2^B}}\left( {{R_{22}}\sqrt {{\Gamma _{A2}}} {e^{j{k_0}\left| {x - x_2^A} \right|}} + {R_{32}}\sqrt {{\Gamma _{A3}}} {e^{j{k_0}\left| {x - x_1^A} \right|}}} \right);
    \end{dmath}
\begin{dmath}
\left\langle {{x_B}} \right.\left| {{\Psi _B}} \right\rangle  = {e^{j{k_0}x}} - j\sqrt {{\Gamma _{B1}}} {e^{j{k_0}x_1^B}}\left( {{R_{11}}\sqrt {{\Gamma _{B2}}} {e^{j{k_0}\left| {x - x_1^B} \right|}} + {R_{21}}\sqrt {{\Gamma _{B1}}} {e^{j{k_0}\left| {x - x_2^B} \right|}}} \right) - j\sqrt {{\Gamma _{B2}}} {e^{j{k_0}x_2^B}}\left( {{R_{12}}\sqrt {{\Gamma _{B1}}} {e^{j{k_0}\left| {x - x_1^B} \right|}} + {R_{22}}\sqrt {{\Gamma _{B2}}} {e^{j{k_0}\left| {x - x_2^B} \right|}}} \right);
    \end{dmath}
\begin{dmath}
\left\langle {{x_C}} \right.\left| {{\Psi _B}} \right\rangle  =  - j\sqrt {{\Gamma _{B1}}} {e^{j{k_0}x_1^B}}\left( {{R_{11}}\sqrt {{\Gamma _{C1}}} {e^{j{k_0}\left| {x - x_2^C} \right|}} + {R_{41}}\sqrt {{\Gamma _{C4}}} {e^{j{k_0}\left| {x - x_1^C} \right|}}} \right) - j\sqrt {{\Gamma _{B2}}} {e^{j{k_0}x_2^B}}\left( {{R_{12}}\sqrt {{\Gamma _{C1}}} {e^{j{k_0}\left| {x - x_2^C} \right|}} + {R_{42}}\sqrt {{\Gamma _{C4}}} {e^{j{k_0}\left| {x - x_1^C} \right|}}} \right);
    \end{dmath}
\begin{dmath}
\left\langle {{x_A}} \right.\left| {{\Psi _C}} \right\rangle  =  - j\sqrt {{\Gamma _{C1}}} {e^{j{k_0}x_2^C}}\left( {{R_{21}}\sqrt {{\Gamma _{A2}}} {e^{j{k_0}\left| {x - x_2^A} \right|}} + {R_{31}}\sqrt {{\Gamma _{A3}}} {e^{j{k_0}\left| {x - x_1^A} \right|}}} \right) - j\sqrt {{\Gamma _{C4}}} {e^{j{k_0}x_1^C}}\left( {{R_{24}}\sqrt {{\Gamma _{A2}}} {e^{j{k_0}\left| {x - x_2^A} \right|}} + {R_{34}}\sqrt {{\Gamma _{A3}}} {e^{j{k_0}\left| {x - x_1^A} \right|}}} \right);
    \end{dmath}
\begin{dmath}
\left\langle {{x_B}} \right.\left| {{\Psi _C}} \right\rangle  =  - j\sqrt {{\Gamma _{C1}}} {e^{j{k_0}x_2^C}}\left( {{R_{11}}\sqrt {{\Gamma _{B2}}} {e^{j{k_0}\left| {x - x_1^B} \right|}} + {R_{21}}\sqrt {{\Gamma _{B1}}} {e^{j{k_0}\left| {x - x_2^B} \right|}}} \right) - j\sqrt {{\Gamma _{B2}}} {e^{j{k_0}x_2^B}}\left( {{R_{14}}\sqrt {{\Gamma _{B1}}} {e^{j{k_0}\left| {x - x_1^B} \right|}} + {R_{24}}\sqrt {{\Gamma _{B2}}} {e^{j{k_0}\left| {x - x_2^B} \right|}}} \right);
    \end{dmath}
\begin{dmath}
\left\langle {{x_C}} \right.\left| {{\Psi _C}} \right\rangle  = {e^{j{k_0}x}} - j\sqrt {{\Gamma _{C1}}} {e^{j{k_0}x_2^C}}\left( {{R_{11}}\sqrt {{\Gamma _{C1}}} {e^{j{k_0}\left| {x - x_2^C} \right|}} + {R_{41}}\sqrt {{\Gamma _{C4}}} {e^{j{k_0}\left| {x - x_1^C} \right|}}} \right) - j\sqrt {{\Gamma _{C4}}} {e^{j{k_0}x_1^C}}\left( {{R_{14}}\sqrt {{\Gamma _{C1}}} {e^{j{k_0}\left| {x - x_2^C} \right|}} + {R_{44}}\sqrt {{\Gamma _{C4}}} {e^{j{k_0}\left| {x - x_1^C} \right|}}} \right);
    \end{dmath}
    \end{dgroup*}
\end{subequations}
\end{widetext}
where we have introduced waveguide photon’s state in a configuration basis and used $\left\langle {{x_n}} \right.\left| {{k_m}} \right\rangle  = {\delta _{mn}}{e^{j{k_m}{x_n}}}$  , which is raised from definitions of photon states, for example $\left| {{k_A}} \right\rangle  = a_k^\dag \left| 0 \right\rangle $ and $\left| {{x_A}} \right\rangle  = \sum\limits_k {a_k^\dag {e^{ik{x_A}}}\left| 0 \right\rangle }  \otimes \left| G \right\rangle $. Henceforward we will omit these indices of $k$ and $x$ to not overload the equations, and we will comment it when it is needed.
Let’s describe meaning of these solutions for one of the initial state $\left| A \right\rangle $  . The equation (\ref{psi_aa}) describes wave function, which can be used to detect the photon in $A$ waveguide (independently of its direction), if the initially photon was sent to this waveguide $A$. The first term just refers to a wave of initially sent photon. If we suppose, that there are no any interaction between this photon and qubits Q2 and Q3 (${\Gamma _{A2}} = {\Gamma _{A3}} = 0$ ), we simply get $\left\langle {{x_A}} \right.\left| {{\Psi _A}} \right\rangle  = {e^{j{k_0}x}}$ . Note that modulus of this function gives a unity, which is absolutely understandable from a physical point of view. In this case, other outcomes have zero probabilities. Equations (\ref{psi_ab}) and (\ref{psi_ac}) don’t contain the ${e^{j{k_0}x}}$  term, because initially there is no any incident photon. The moduli of coordinate differences simply arise from the integrals, as for example it was shown \cite{22}. So, these moduli have a clear meaning, because their sign defines transmission and reflection coefficients. For example, if we are interested in reflection coefficient of (\ref{psi_aa}) (or in other words, in probability to find photon in the left side of qubit 3 in $A$ waveguide), we just set the following assumptions:
\begin{equation}
x < x_2^A \Rightarrow \left\{ \begin{array}{l}
{e^{j{k_0}\left| {x - x_2^A} \right|}} = {e^{ - j{k_0}x}} \cdot {e^{j{k_0}x_2^A}};\\
{e^{j{k_0}\left| {x - x_1^A} \right|}} = {e^{ - j{k_0}x}} \cdot {e^{j{k_0}x_1^A}};
\end{array} \right.
\label{mod_1}
\end{equation}
where ${e^{ - j{k_0}x}}$   term describes the wave propagating in a left direction. According to this disclosures (\ref{mod_1}) we write the wave function as follows:
\begin{widetext}
\begin{dmath*}
\left\langle {{x_A}} \right.\left| {{\Psi _A}} \right\rangle  = {e^{j{k_0}x}} - j\sqrt {{\Gamma _{A2}}} {e^{ - j{k_0}x}}\left( {{R_{22}}\sqrt {{\Gamma _{A2}}}  \cdot {e^{2j{k_0}x_2^A}} + {R_{32}}\sqrt {{\Gamma _{A3}}}  \cdot {e^{j{k_0}\left( {x_2^A + x_1^A} \right)}}} \right)\\ 
- j\sqrt {{\Gamma _{A3}}} {e^{ - j{k_0}x}}\left( {{R_{23}}\sqrt {{\Gamma _{A2}}}  \cdot {e^{j{k_0}\left( {x_2^A + x_1^A} \right)}} + {R_{33}}\sqrt {{\Gamma _{A3}}}  \cdot {e^{2j{k_0}x_1^A}}} \right);
\end{dmath*}
\end{widetext}
 then by natural defining reflection coefficient as ratio of counter propagating wave to direct propagating wave we get reflection coefficient as:
\begin{dmath}\label{r_AA}
{r_{AA}} =  - j\sqrt {{\Gamma _{A2}}} \left( {{R_{22}}\sqrt {{\Gamma _{A2}}} {e^{2j{k_0}x_2^A}} + {R_{32}}\sqrt {{\Gamma _{A3}}} {e^{j{k_0}\left( {x_2^A + x_1^A} \right)}}} \right)\\
 - j\sqrt {{\Gamma _{A3}}} \left( {{R_{23}}\sqrt {{\Gamma _{A2}}} {e^{j{k_0}\left( {x_2^A + x_1^A} \right)}} + {R_{33}}\sqrt {{\Gamma _{A3}}} {e^{2j{k_0}x_1^A}}} \right).
\end{dmath}

Or, controversially we can set  $x > x_1^A$, and get the transmission coefficient in the waveguide $A$:
\begin{equation}\label{t_AA}
\begin{split}
{t_{AA}} = 1 - j\sqrt {{\Gamma _{A2}}} \left( {{R_{22}}\sqrt {{\Gamma _{A2}}}  + {R_{32}}\sqrt {{\Gamma _{A3}}} {e^{j{k_0}\left( {x_2^A - x_1^A} \right)}}} \right)\\
 - j\sqrt {{\Gamma _{A3}}} \left( {{R_{23}}\sqrt {{\Gamma _{A2}}} {e^{j{k_0}\left( {x_1^A - x_2^A} \right)}} + {R_{33}}\sqrt {{\Gamma _{A3}}} } \right).
\end{split}
\end{equation}
These equations (\ref{r_AA})-(\ref{t_AA}) can be transformed to well-known results in the limit cases: 1) only coupling Q3 to $A$ waveguide is non-zero 2) only couplings of Q3 and Q2 to $A$ waveguide are non-zero. In the first limit we get the transmission and reflection coefficients demonstrated experimentally by Astafiev et al \cite{28}:
 \begin{align*}
{r_{AA}} &= \frac{{ - j{\Gamma _{A3}}{e^{2j{k_0}x_1^A}}}}{{\omega  - {\Omega _3} + j{\Gamma _{A3}}}};\\
{t_{AA}} &= 1 - j{\Gamma _{A3}}\frac{1}{{\omega  - {\Omega _3} + j{\Gamma _{A3}}}};
\end{align*}
because at this case the effective matrix becomes diagonal. Here it is obvious, that at the frequency equal to ${\Omega _3}$  the reflection coefficient is equal to 1, and this fact is caused by the interference of the reflected and initial wave functions. From this point of view, the functionality of  qubits Q3 and Q4 is quite clear, they serve as the additional scatterers  to create appropriate interference conditions for the routing, while  coupling between the waveguides can be provided by qubits Q1 and Q2. The second limit leads to the results presented \cite{22,29,30}, saving all distance and frequency dependencies.
The conditions below define the reflection and transmission, when photon after the scattering could be found at $B$ or $C$ waveguide, respectively:
\begin{equation}
\begin{array}{l}
x < x_1^B\,\,\,\,and\,\,\,\,x > x_2^B\\
x < x_1^C\,\,\,\,and\,\,\,\,x > x_2^C
\end{array}
\label{mod_2}
\end{equation}
We have omitted the indeces of $x$ in the exponent of photon wave function ${e^{j{k_m}{x_n}}}$, but if consider these indeces in scattering problem we get the relation like $\frac{{{e^{j{k_0}{x_{\left( {B,C} \right)}}}}}}{{{e^{j{k_0}{x_{\left( A \right)}}}}}} \equiv {e^{j\phi }}$  (for photon initially sent to $A$ waveguide we should normalize functions (\ref{psi_ab}) and (\ref{psi_ac}) to initial wave ${e^{j{k_0}{x_{\left( A \right)}}}}$ ). This element simply adds some phase shift $\phi$ for scattered photon, and this fact is natural due to some phase incursion while travelling in “perpendicular” to the waveguide direction. At the next section we introduce conditions for one-dimensionality, based on neglecting this phase incursions. But in general, all equations (\ref{psi_f}) consider these phase additions.
Henceforward we will specify ${r_{fin - in}}$  as the reflection coefficient, describing the probability to find the photon in the final (index $fin$) waveguide’ left side (because we chose a right direction as positive x-axis), when initially photon was sent in the initial waveguide(index $in$). These initial and final waveguides refer to $A, B, C$ waveguides, which could contain photon before and after scattering as well. The transmission coefficient ${t_{fin - in}}$  is specified with the same sense. In some simplification, the difference between the reflection and transmission is just question about left or right sides at Fig. \ref{fig1}. One can see, that these coefficients (\ref{r_AA}) and (\ref{t_AA}) depend on the distance between qubits Q3 and Q2, and it results from the retardation effect (it was described in \cite{22}). However, the more intriguing fact is that they both depend now on the distances between other qubits because each term contains this information from the effective Hamiltonian’s inversion (\ref{h_eff}). It is a direct manifestation of absolutely quantum interferences between different wave functions, which are considered in all orders of interaction in a frame of the method.
It can be shown by a direct substitution, that for each initial state manifold, the normalization condition is always satisfied:
\begin{equation}
\sum\limits_{J = A,B,C} {{{\left| {{t_{J - in}}} \right|}^2} + {{\left| {{r_{J - in}}} \right|}^2}}  = 1
\end{equation}
The calculations of all transmission and reflection amplitudes are presented in Appendix A.
It is worth noting that all solutions (\ref{psi_f}) are obtained without any simplification in a sense of distances. All coefficients could be defined simply, but when calculating moduli we get expressions depending on distances $x_i^J - x_j^M\left( {J \ne M} \right)$. This fact allows us to introduce one-dimensionality because we don’t need exact coordinates of each scatterer. So here we have come to the point of simplification to one-dimensionality condition.

\section{Conditions of one-dimensionality}
\begin{figure}[h!]
\includegraphics[width=\linewidth]{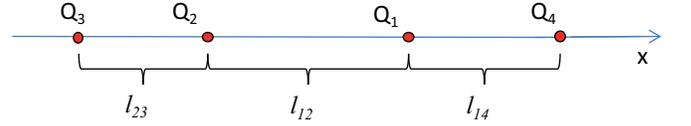}
  \centering
  
  \caption{One-dimensional presentation of the system}
  \label{fig2}
\end{figure}

Here we start our reasoning from definition of distances between points at Fig. \ref{fig1}, as it is shown in Table \ref{table:dist}.
\begin{table}
\caption{Definitions of distances between qubits}
\label{table:dist}
\begin{tabular}{|l|c|}
  \hline
  ${d_{{Q_3} - {n_1}}}$ & distance between $Q_3$ and point $n_1$  \\
  \hline 
 ${d_{{Q_2} - {n_1}}}$ & distance between $Q_2 $ and point $n_1$  \\
  \hline 
 ${d_{{Q_2} - {n_2}}}$ & distance between $Q_2$ and point $n_2$  \\
  \hline 
 ${d_{{n_2} - {n_3}}}$ & distance between points $n_2$ and $n_3$  \\
  \hline 
 ${d_{{Q_1} - {n_3}}}$ & distance between $Q_1$ and point $n_3$  \\
  \hline 
 ${d_{{Q_1} - {n_4}}}$ & distance between $Q_1$ and point $n_4$  \\
  \hline 
 ${d_{{Q_4} - {n_4}}}$ & distance between $Q_4$ and point $n_4$  \\
  \hline 
\end{tabular}
\end{table}
Just for the sake of simplicity let set ${d_{{Q_2} - {n_1}}},{d_{{Q_2} - {n_2}}},{d_{{Q_1} - {n_3}}},{d_{{Q_1} - {n_4}}}$   to be equal each other and assign this distance as $\delta$. In this case we can define distances ${l_{mn}}$  between $m^{th}$ and $n^{th}$ qubits as:
\begin{align*}
\begin{array}{l}
{l_{12} }= {d_{{n_2} - {n_3}}} + 2\delta ;\\
{l_{13}} = {d_{{Q_3} - {n_1}}} + {d_{{n_2} - {n_3}}} + 2\delta ;\\
{l_{14}} = {d_{{Q_4} - {n_4}}} + \delta ;\\
{l_{23}} = {d_{{Q_3} - {n_1}}} + \delta ;\\
{l_{24}} = {d_{{n_2} - {n_3}}} + {d_{{Q_4} - {n_4}}} + 3\delta ;\\
{l_{34}} = {d_{{Q_3} - {n_1}}} + {d_{{n_2} - {n_3}}} + {d_{{Q_4} - {n_4}}} + 4\delta .
\end{array}
\end{align*}
One-dimensionality is introduced as the following conditions:
\begin{equation}\label{1Dcond}
4\delta  <  < {d_{{Q_3} - {n_1}}},{d_{{n_2} - {n_3}}},{d_{{Q_4} - {n_4}}}
\end{equation}
These conditions could be easily realized in solid-state superconducting chips, because distances like a $\delta$  usually should be several hundreds of micrometers, while other lengths are typically in a centimeter range. The inequality (\ref{1Dcond}) leads us to the following relations:
 \begin{align*}
\begin{array}{l}
{l_{12}} \approx {d_{{n_2} - {n_3}}};\\
{l_{13}} \approx {d_{{Q_3} - {n_1}}} + {d_{{n_2} - {n_3}}};\\
{l_{14}} \approx {d_{{Q_4} - {n_4}}};\\
{l_{23}} \approx {d_{{Q_3} - {n_1}}};\\
{l_{24}} \approx {d_{{n_2} - {n_3}}} + {d_{{Q_4} - {n_4}}};\\
{l_{34}} \approx {d_{{Q_3} - {n_1}}} + {d_{{n_2} - {n_3}}} + {d_{{Q_4} - {n_4}}}.
\end{array}
\end{align*}
It directly leads to that the scheme presented at Fig \ref{fig1}, could be transformed in a pseudo-one dimensional axis as shown at Fig \ref{fig2}.
We can specify the zero coordinate point of this pseudo-axis as the middle point between $Q_2$ and $Q_1$. In this case we redefine the qubits’s coordinates as follows:
 \begin{equation}
\begin{array}{l}
x_1^A =  - {l_{23}} - \frac{1}{2}{l_{12}};\\
x_2^A = x_2^B =  - \frac{1}{2}{l_{12}};\\
x_1^B = x_2^C = \frac{1}{2}{l_{12}};\\
x_1^C = {l_{14}} + \frac{1}{2}{l_{12}};
\end{array}
\end{equation}
These equalities are used to disclose the modulus signs in equations (\ref{psi_f}). But we should  again mention that the final results don’t depend on the zero coordinate choice.

\section{Simulations and functionality of the router}

\begin{figure}[h]
\begin{subfigure}[b]{0.5\textwidth}
\includegraphics[width=\textwidth]{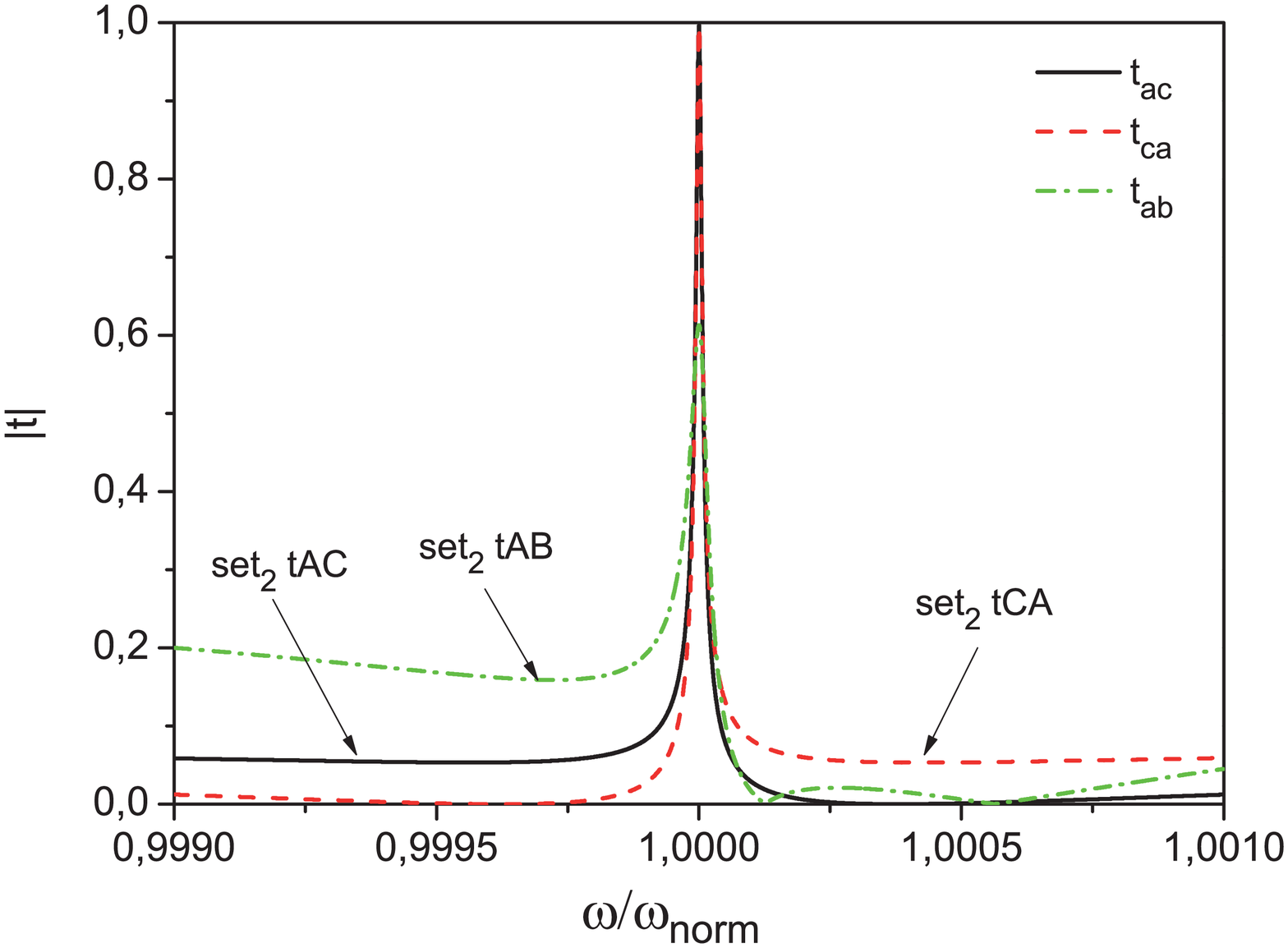}
    \caption{}
\label{fig3a}
\end{subfigure}
\hfill 
\begin{subfigure}[b]{0.5\textwidth}
\includegraphics[width=\textwidth]{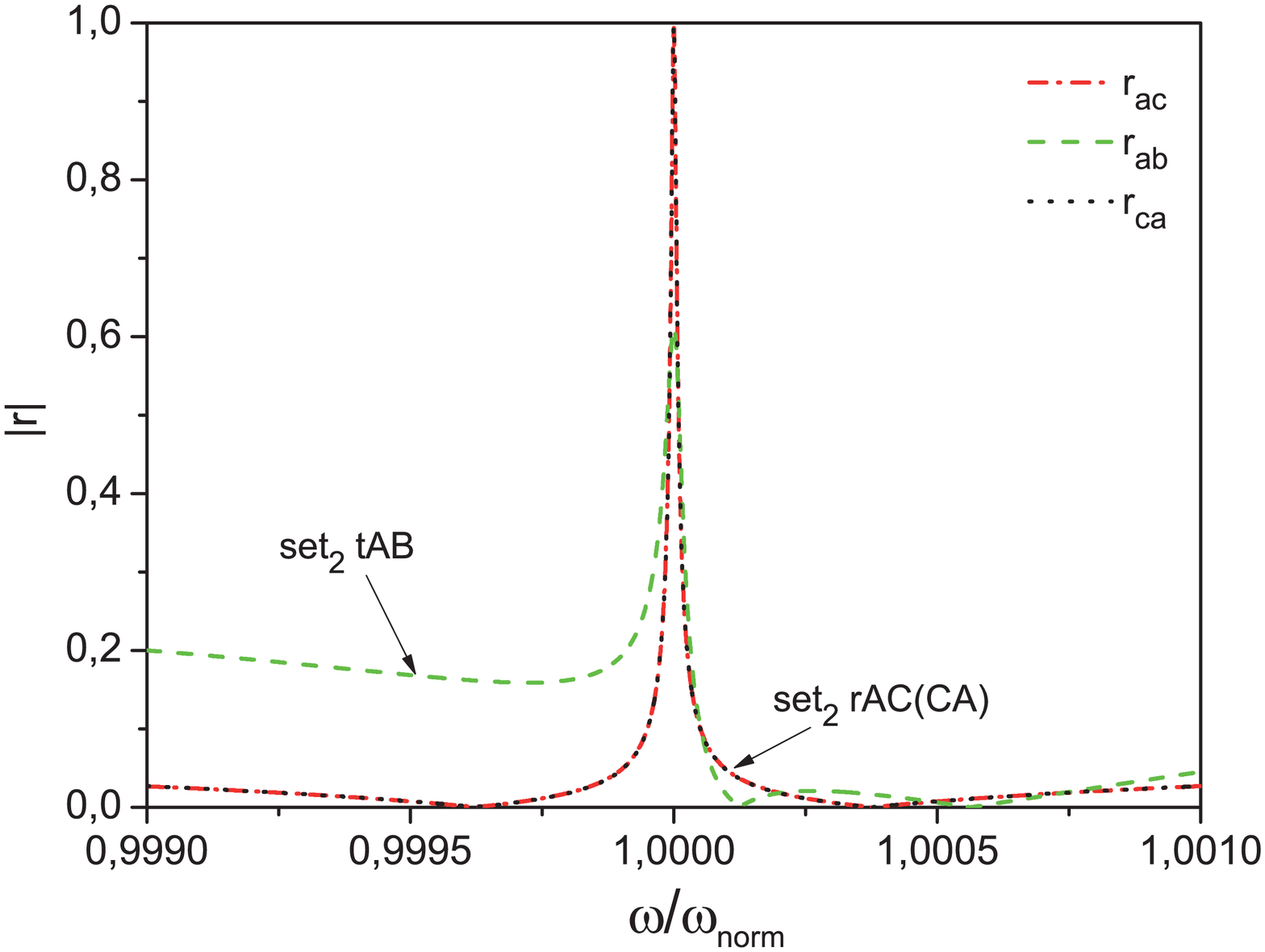}
    \caption{}
\label{fig3b}
\end{subfigure}
\caption{Transmission and reflection coefficients for the router with the following parameters: $\beta=3.4$, $L_{side}=0.028$, $L_{12}=2\pi$ , $\Theta_N$ is set to $5 GHz$, $\Gamma_{central}=10 MHz$ . The sets are described at Table \ref{table:set_4} }
\label{fig3}
\end{figure}

\begin{figure*}
\centering
\begin{subfigure}[b]{0.48\textwidth}
\centering
\includegraphics[width=\textwidth]{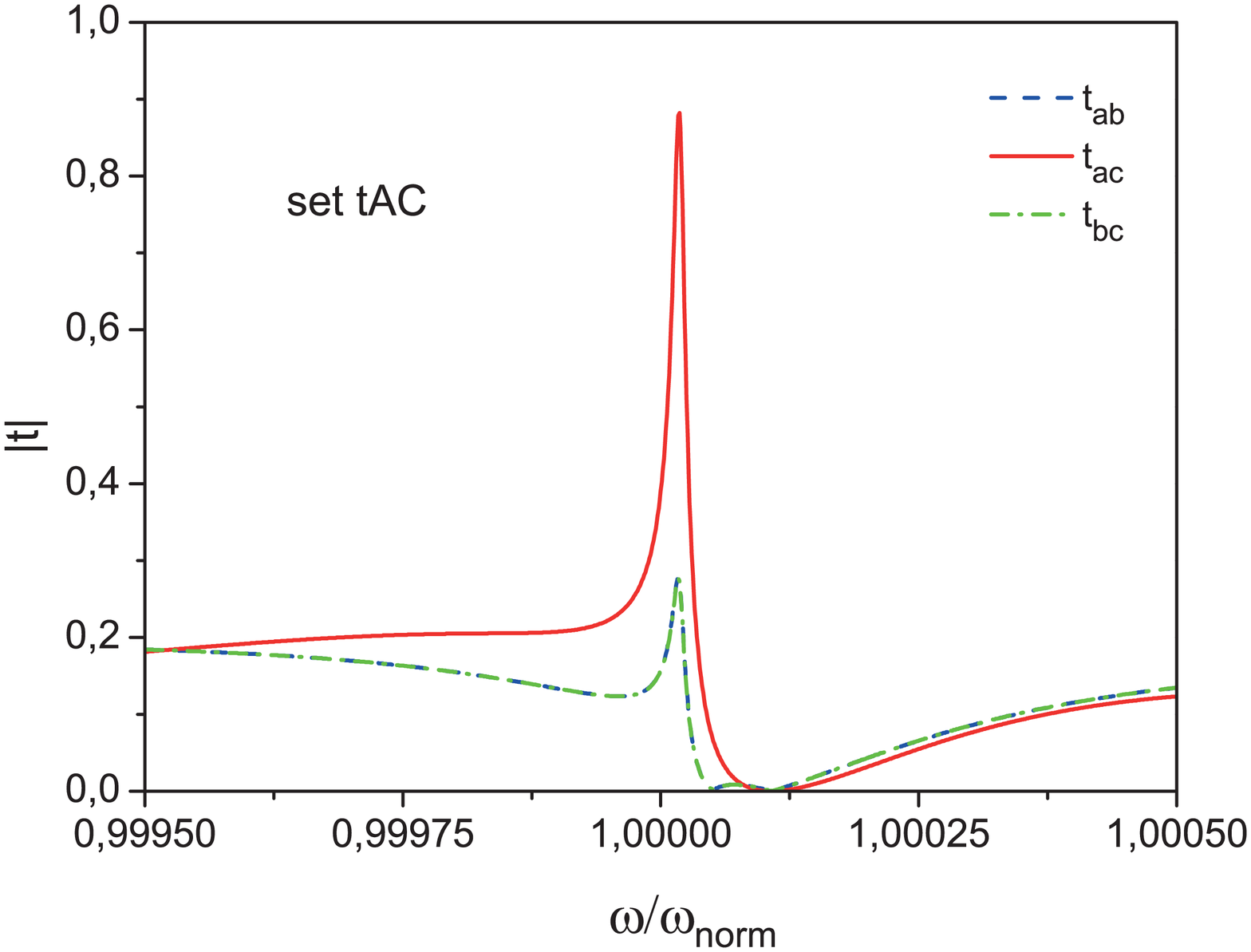}
\caption{}%
\label{fig4a}
\end{subfigure}
\hfill
\begin{subfigure}[b]{0.48\textwidth}
\centering
\includegraphics[width=\textwidth]{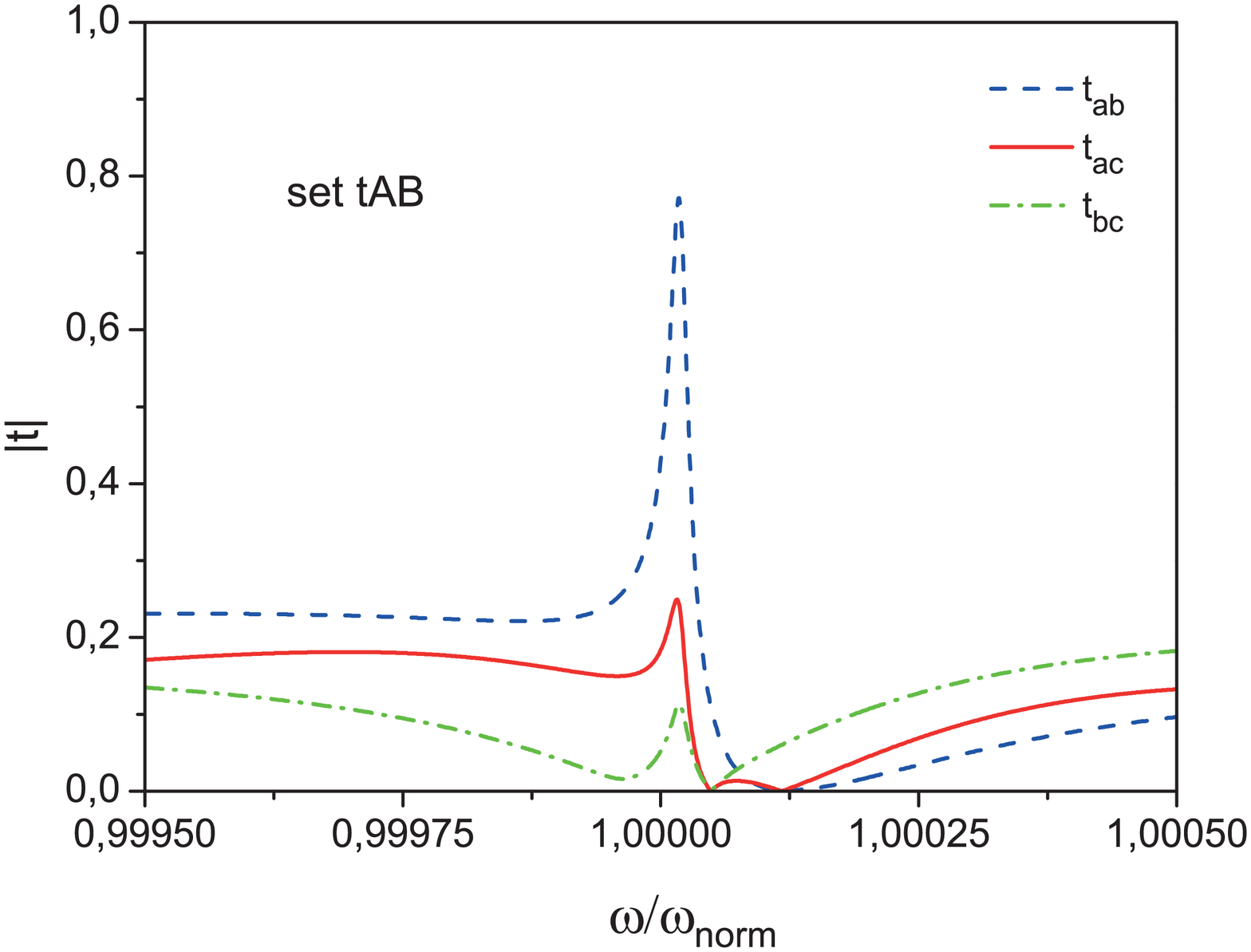}
\caption{}%
\label{fig4b}
\end{subfigure}
\vskip\baselineskip
\begin{subfigure}[b]{0.48\textwidth}
\centering
\includegraphics[width=\textwidth]{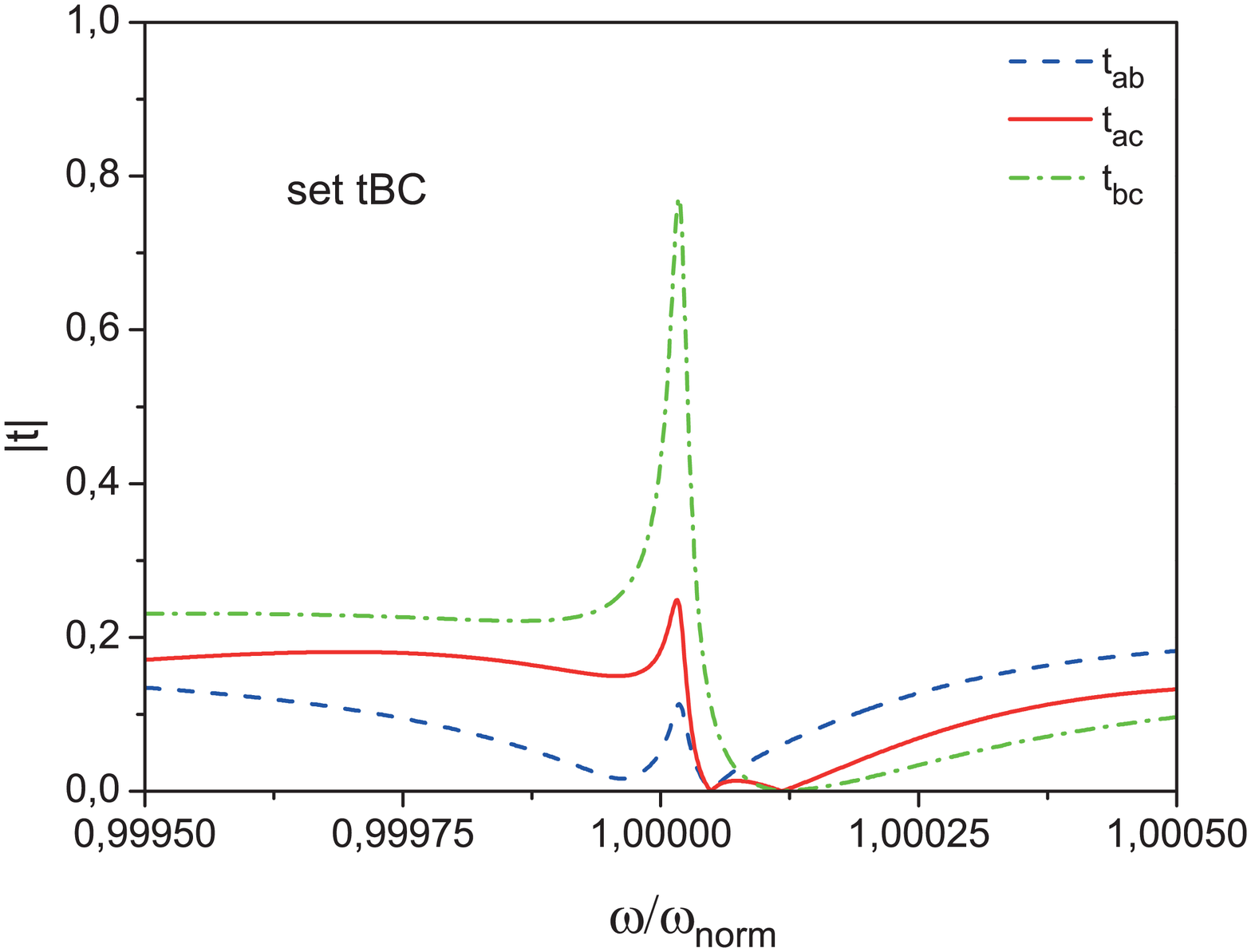}
\caption{}%
\label{fig4c}
\end{subfigure}
\quad
\begin{subfigure}[b]{0.48\textwidth}
\centering
\includegraphics[width=\textwidth]{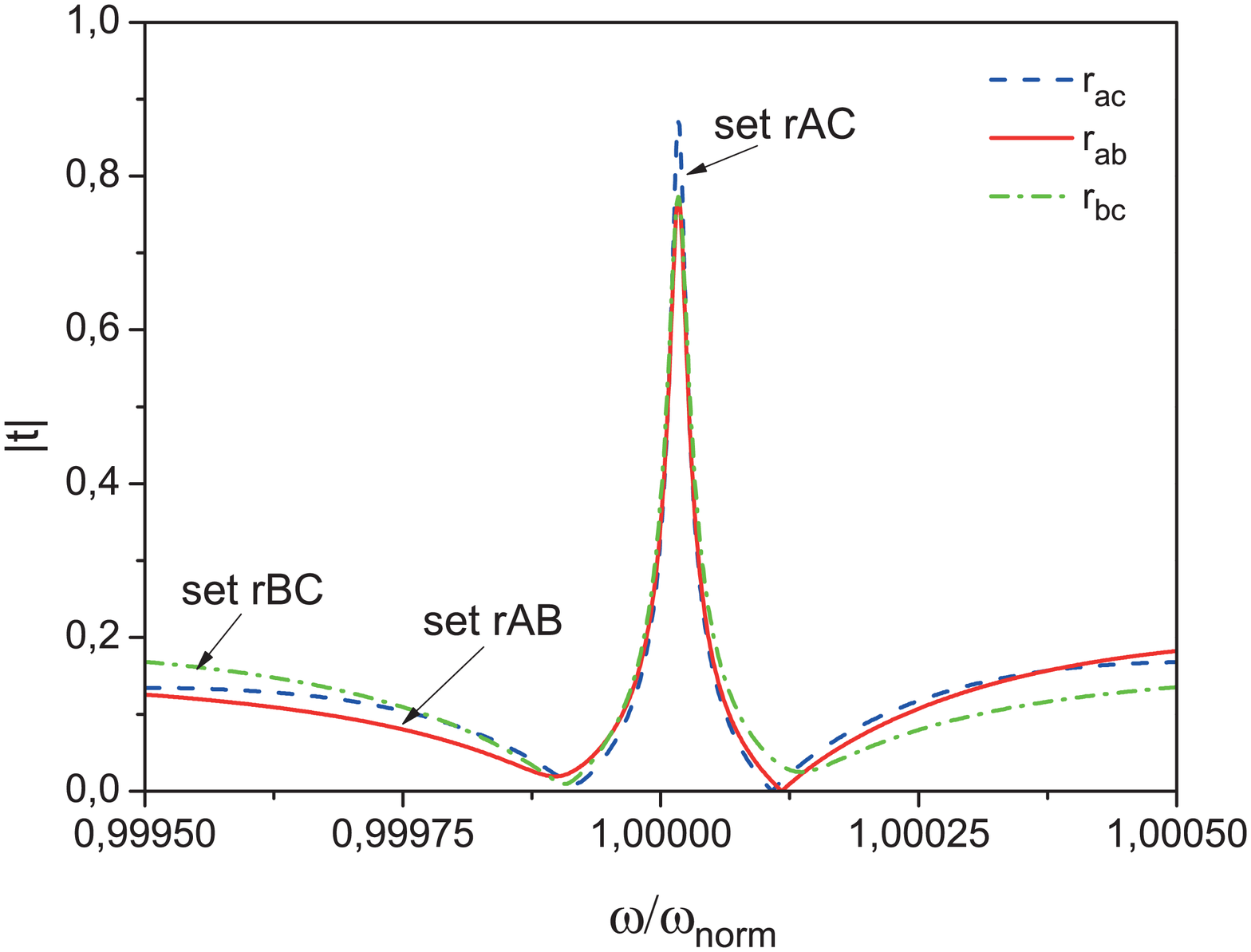}
\caption{}%
\label{fig4d}
\end{subfigure}
\caption{Tunability of routing by setting qubits’ excitations frequencies (set fIJ refers to combinations of to provide maximum transmission(f=t) or reflection coefficients(f=r) between waveguides $I$ and $J$ ($I,J=A,B,C$ ). Parameters are the following: $\beta=0.2$, $L_{side}=\pi/30$, $L_{12}=0.01\pi$ , $\Theta_N$ is set to $5 GHz$, $\Gamma_{central}=10 MHz$ . The sets are described at Table \ref{table:set_6} }
\label{fig4}
\end{figure*}

Because the router has symmetry relative to $B$ waveguide, it is intuiteve to set the following equalities between distances ${l_{23}} = {l_{14}} = {l_{side}}$ and couplings ${\Gamma _{A3}} = {\Gamma _{A2}} = {\Gamma _{C1}} = {\Gamma _{C4}} = {\Gamma _{side}}$
and ${\Gamma _{B1}} = {\Gamma _{B2}} = {\Gamma _{central}}$ . Also it is possible to introduce relations between the central waveguide and side waveguides parameters:
\begin{equation}
\frac{{{\Gamma _{side}}}}{{{\Gamma _{central}}}} = \beta
\end{equation}
The most flexible parameters include the following distances:
\begin{equation}
\begin{array}{l}
\frac{{{\Theta _N}}}{{{\nu _g}}}{l_{side}} = {L_{side}}\\
\frac{{{\Theta _N}}}{{{\nu _g}}}{l_{12}} = {L_{12}}
\end{array}
\end{equation}
where we introduced some fixed normalization frequency ${\Theta _N}$ due to generalizing the distances. This introduction leads to a substituting:
\begin{equation*}
{k_0} = \frac{\omega }{{{\nu _g}}}\frac{{{\Theta _N}}}{{{\Theta _N}}} = \frac{\omega }{{{\Theta _N}}}{k_\Theta }
\end{equation*}
The optimal distances have been found semi numerically by Quasi-Newton methods for extremum search with initial conditions defined by the following assumptions:

-    central waveguide should introduce minimum of phase increasing, because signal between $Q_1$ and $Q_2$ has minimal opportunity to escape in comparison for qubits $Q_3$ and $Q_4$ and we should not provide strong interference conditions at this region(for flexibility)\\
-    the couplings $Q_1$ and $Q_2$ to the central waveguide should be greater than other couplings, because the waveguide provide intermediate interaction between $A$ and $C$ waveguides.\\
Firstly, let us note, that it is possible to find such parameters, which provide near 1 probability to detect photon at each of six router’ outputs. Unfortunately, we have found only such sets of preset parameters (distances between qubits, coupling strengths etc.) which can provide the maximum not for all probabilities (for example, just for two ports of $B$ waveguide), making impossible tuning each probability by  controllable parameters (qubits’ excitation frequency) in situ. It is shown at the Fig.\ref{fig3}
From Fig.\ref{fig3} it is obvious that the probability to move photon inside(outside) the $B$ waveguide cannot be more than 0.6. Moreover such configuration strongly depends on photon frequency, and it was found that it has bandwidth about $60 MHz$ (losing 15 \% of probability), repeated around the frequencies ${\Theta _x} = 2n{\Theta _N},n = 1,2...$ . But at least it can be used as 4 port device.
\begin{table}
\caption{Example sets for tuning the routing as at Fig \ref{fig3}.}
\label{table:set_4}
\resizebox{0.475\textwidth}{!}{
\begin{tabular}{|c|c|c|c|c|}
\hline
& $\Omega_1-\Theta_N$, MHz & $\Omega_2-\Theta_N$, MHz & $\Omega_3-\Theta_N$, MHz & $\Omega_4-\Theta_N$, MHz \\
\hline
$set_2 \; tAC$& $-0.017$ & $-0.024$ & $1.901$ & $1.898$ \\
\hline
$set_2 \; tCA$& $1.884$ & $1.879$ & $-0.0033$ & $-0.0077$ \\
\hline
$set_2 \;tAB$& $2.193$ & $-2.155$ & $2.811$ & $ 0.939$ \\
\hline
$set_2 \; rAC$& $1.883$ & $-0.0076$ & $1.886$ & $-0.0067$ \\
\hline
\end{tabular}
}
\end{table}
Of course, we should define more flexible preset parameters for practical 6 port device. At this section we show how to tune the probabilities of six outputs by controllable parameters. All transmission and reflection coefficients are defined exactly at Appendix A.
At Fig.\ref{fig4} transmission (a, b, c) and reflection(d) coefficients for different combinations of qubits frequencies are shown.
One sees that the router could provide tunable probabilities in a range of 0.8 just by tuning the qubits frequencies (for example by external electrical or magnet field). These sets are listed bellow at the Table \ref{table:set_6}. Also it is worth to mention, that if set all distances to be equal to zero it will be impossible to tune the probabilities more than 0.5. It is indirect manifestation that the routing is based on interference and retardation effect, which are naturally described by non-Hermitian Hamiltonian approach.
\begin{table}
\caption{Example sets for tuning the routing as at Fig \ref{fig4}.}
\label{table:set_6}
\resizebox{0.475\textwidth}{!}{
\begin{tabular}{|c|c|c|c|c|}
\hline
& $\Omega_1-\Theta_N$, MHz & $\Omega_2-\Theta_N$, MHz & $\Omega_3-\Theta_N$, MHz & $\Omega_4-\Theta_N$, MHz \\
\hline
$set \; tAC$& $0.339$ & $0.337$ & $0.53$ & $0.53$ \\
\hline
$set \; tAB$& $1.463$ & $-0.542$ & $0.581$ & $0.239$ \\
\hline
$set \; tBC$& $-0.545$ & $1.465$ & $0.239$ & $ 0.581$ \\
\hline
$set \; rAC$& $0.893$ & $0.334$ & $0.534$ & $-0.048$ \\
\hline
$set \; rAB$& $-0.234$ & $1.13$ & $0.581$ & $0.239$ \\
\hline
$set \; rAC$& $0.086$ & $1.479$ & $0.239$ & $-0.096$ \\
\hline
\end{tabular}
}
\end{table}
    From the Table \ref{table:set_6} it is seen that the minimal difference between qubits’ excitation frequencies is around 200 kHz and this can be easily provided by modern superconducting control schemes.
    We have checked that such sets could be found in a range of photon frequency from 2 GHz to 15 GHz (with fixed normalization ${\Theta _N} = 5\, GHz$, or put it another way with fixed distances ) without any significant loss the maximum probability amplitude.
    The easiest control could be provided for scatterings without waveguide changing (photon stays at the same waveguide, in which it has been sent), and it is naturally obvious. For example when qubits are not in the resonance with the photon, the last simply goes through the system without scatterings. One important thing should be mentioned about the relaxation rates of qubits. Of course, existing of other quantum channels should decrease the probabilities to found photon at the waveguides, but it is enough to provide coupling relation $\frac{{{\Gamma _{side}}}}{\gamma } \ge 10$ where $\gamma$ is the maximum relaxation rate. This relation has been checked by numerical simulations and substituting ${\Omega _i} \to {\Omega _i} - j\gamma $ , because it is legal for single photon scattering and in the absence of a common decay channel.
A follow up study should include expanded space of states, which are considered in a frame of the method. For example, it could be realized in a problem of many photon scattering or if we expand internal states’ basis with two or more qubit’ excited states (or similarly two photon states in waveguides). At this case, we get opportunity to create not only superpositions states after scattering like $\alpha \left| {{k_M}} \right\rangle \otimes \left| G \right\rangle + \beta \left| {{k_N}} \right\rangle \otimes \left| G \right\rangle $ and tune them like it is shown here, where $M, N=A,B,C$, but additionally some Bell states (for example like a $\alpha \left| {{k_M}} \right\rangle \otimes \left| {{g_1},{g_2},{e_3},{g_4}} \right\rangle + \beta \left| {{k_N}} \right\rangle \otimes \left| {{g_1},{g_2},{g_3},{e_4}} \right\rangle $ ) could be realized, if only one photon will escape the system after two-photon scattering.
\section*{Conclusion}
At this work we offered the design of single photon router, and presented quantum mechanical method for the calculation of its performance. The obtained equations are quite general, taking into account a non-uniformity of artificial atoms parameters as well as all distances between scatterers. Corresponding limiting cases demonstrate a validity of our approach. For single photon we obtained analytical expressions for probabilities to detect it at each waveguide.
We have shown how this device can be used in a wide- and narrowband regimes. For the narrowband regime, we reduce the number of operating ports to 4 ports (the central waveguide provides interaction between two others). We have shown, that the probability to detect a photon in each port can be set to near unity by an appropriate tuning of the qubits' excitation frequency. For the wideband regime, we numerically found optimal parameters, which allow to tune a routing quality more than 0.75 for each of 6 ports.
\section*{Acknowledgement}
A.N. thanks IPHT for the hospitality during the project 3.12764.2018/12.2 of Russian Ministry of Science and Higher Education and DAAD. Ya. S. G. acknowledges the support from Russian Ministry of Education and Science under the project
3.4571.2017/6.7.
\appendix{}
\section{}
To define scattering parameters, we should correctly disclose moduli signs in (\ref{psi_f}) equations. We do it by definitions (\ref{mod_1}) and (\ref{mod_2}) and get set of 18 scattering amplitudes.
For initially photon propagating in the $A$ waveguide we have three transmission amplitudes:
\begin{dmath}
{t_{AA}} = 1 - j\sqrt {{\Gamma _{A2}}} \left( {{R_{22}}\sqrt {{\Gamma _{A2}}}  + {R_{32}}\sqrt {{\Gamma _{A3}}} {e^{j{k_0}{l_{23}}}}} \right)\\ - j\sqrt {{\Gamma _{A3}}} \left( {{R_{23}}\sqrt {{\Gamma _{A2}}} {e^{ - j{k_0}{l_{23}}}} + {R_{33}}\sqrt {{\Gamma _{A3}}} } \right)
\end{dmath}

\begin{dmath}
{t_{AB}} =  - j\sqrt {{\Gamma _{A2}}} \left( {{R_{12}}\sqrt {{\Gamma _{B1}}} {e^{ - j{k_0}{l_{12}}}} + {R_{22}}\sqrt {{\Gamma _{B2}}} } \right) - j\sqrt {{\Gamma _{A3}}} \left( {{R_{13}}\sqrt {{\Gamma _{B1}}} {e^{ - j{k_0}\left( {{l_{12}} + {l_{23}}} \right)}} + {R_{23}}\sqrt {{\Gamma _{B2}}} {e^{ - j{k_0}{l_{23}}}}} \right)
\end{dmath}

\begin{dmath}
{t_{AC}} =  - j\sqrt {{\Gamma _{A2}}} \left( \begin{array}{l}
{R_{12}}\sqrt {{\Gamma _{C1}}} {e^{ - j{k_0}{l_{12}}}}\\
 + {R_{42}}\sqrt {{\Gamma _{C4}}} {e^{ - j{k_0}\left( {{l_{12}} + {l_{14}}} \right)}}
\end{array} \right) - j\sqrt {{\Gamma _{A3}}} \left( \begin{array}{l}
{R_{13}}\sqrt {{\Gamma _{C1}}} {e^{ - j{k_0}\left( {{l_{23}} + {l_{12}}} \right)}}\\
 + {R_{43}}\sqrt {{\Gamma _{C4}}} {e^{ - j{k_0}\left( {{l_{12}} + {l_{23}} + {l_{14}}} \right)}}
\end{array} \right)
\end{dmath}
and three reflection amplitudes:
\begin{dmath}
{r_{AA}} =  - j\sqrt {{\Gamma _{A2}}} \left( \begin{array}{l}
{R_{22}}\sqrt {{\Gamma _{A2}}} {e^{ - j{k_0}{l_{12}}}}\\
 + {R_{32}}\sqrt {{\Gamma _{A3}}} {e^{ - j{k_0}\left( {{l_{12}} + {l_{23}}} \right)}}
\end{array} \right) - j\sqrt {{\Gamma _{A3}}} \left( \begin{array}{l}
{R_{23}}\sqrt {{\Gamma _{A2}}} {e^{ - j{k_0}\left( {{l_{12}} + {l_{23}}} \right)}}\\
 + {R_{33}}\sqrt {{\Gamma _{A3}}} {e^{ - j{k_0}\left( {{l_{12}} + 2{l_{23}}} \right)}}
\end{array} \right)
\end{dmath}
\begin{dmath}
{r_{AB}} =  - j\sqrt {{\Gamma _{A2}}} \left( {{R_{12}}\sqrt {{\Gamma _{B1}}}  + {R_{22}}\sqrt {{\Gamma _{B2}}} {e^{ - j{k_0}{l_{12}}}}} \right) - j\sqrt {{\Gamma _{A3}}} \left( {{R_{13}}\sqrt {{\Gamma _{B1}}} {e^{ - j{k_0}{l_{23}}}} + {R_{23}}\sqrt {{\Gamma _{B2}}} {e^{ - j{k_0}\left( {{l_{12}} + {l_{23}}} \right)}}} \right)
\end{dmath}
\begin{dmath}
{r_{AC}} =  - j\sqrt {{\Gamma _{A2}}} \left( {{R_{12}}\sqrt {{\Gamma _{C1}}}  + {R_{42}}\sqrt {{\Gamma _{C4}}} {e^{j{k_0}{l_{14}}}}} \right) - j\sqrt {{\Gamma _{A3}}} \left( {{R_{13}}\sqrt {{\Gamma _{C1}}} {e^{ - j{k_0}{l_{23}}}} + {R_{43}}\sqrt {{\Gamma _{C4}}} {e^{ - j{k_0}\left( {{l_{23}} - {l_{14}}} \right)}}} \right)
\end{dmath}
Similarly, for initially photon propagating in the $B$ waveguide, there are three transmission amplitudes:
\begin{dmath}
{t_{BA}} =  - j\sqrt {{\Gamma _{B1}}} \left( {{R_{21}}\sqrt {{\Gamma _{A2}}} {e^{j{k_0}{l_{12}}}} + {R_{31}}\sqrt {{\Gamma _{A3}}} {e^{j{k_0}\left( {{l_{12}} + {l_{23}}} \right)}}} \right) - j\sqrt {{\Gamma _{B2}}} \left( {{R_{22}}\sqrt {{\Gamma _{A2}}}  + {R_{32}}\sqrt {{\Gamma _{A3}}} {e^{j{k_0}{l_{23}}}}} \right)
\end{dmath}
\begin{dmath}
{t_{BB}} = 1 - j\sqrt {{\Gamma _{B1}}} \left( {{R_{11}}\sqrt {{\Gamma _{B1}}}  + {R_{21}}\sqrt {{\Gamma _{B2}}} {e^{j{k_0}{l_{12}}}}} \right) - j\sqrt {{\Gamma _{B2}}} \left( {{R_{12}}\sqrt {{\Gamma _{B1}}} {e^{ - j{k_0}{l_{12}}}} + {R_{22}}\sqrt {{\Gamma _{B2}}} } \right)
\end{dmath}
\begin{dmath}
{t_{BC}} =  - j\sqrt {{\Gamma _{B1}}} \left( {{R_{11}}\sqrt {{\Gamma _{C1}}}  + {R_{41}}\sqrt {{\Gamma _{C4}}} {e^{ - j{k_0}{l_{14}}}}} \right) - j\sqrt {{\Gamma _{B2}}} \left( {{R_{12}}\sqrt {{\Gamma _{C1}}} {e^{ - j{k_0}{l_{12}}}} + {R_{42}}\sqrt {{\Gamma _{C4}}} {e^{ - j{k_0}\left( {{l_{12}} + {l_{14}}} \right)}}} \right)
\end{dmath}
and three reflection amplitudes:
\begin{dmath}
{r_{BA}} =  - j\sqrt {{\Gamma _{B1}}} \left( {{R_{21}}\sqrt {{\Gamma _{A2}}}  + {R_{31}}\sqrt {{\Gamma _{A3}}} {e^{ - j{k_0}{l_{23}}}}} \right) - j\sqrt {{\Gamma _{B2}}} \left( {{R_{22}}{{\sqrt {{\Gamma _{A2}}} }^{ - j{k_0}{l_{12}}}} + {R_{32}}\sqrt {{\Gamma _{A3}}} {e^{ - j{k_0}\left( {{l_{23}} + {l_{12}}} \right)}}} \right)
\end{dmath}
\begin{dmath}
{r_{BB}} =  - j\sqrt {{\Gamma _{B1}}} \left( {{R_{11}}\sqrt {{\Gamma _{B1}}} {e^{j{k_0}{l_{12}}}} + {R_{21}}\sqrt {{\Gamma _{B2}}} } \right) - j\sqrt {{\Gamma _{B2}}} \left( {{R_{12}}\sqrt {{\Gamma _{B1}}}  + {R_{22}}\sqrt {{\Gamma _{B2}}} {e^{ - j{k_0}{l_{12}}}}} \right)
\end{dmath}
\begin{dmath}
{r_{BC}} =  - j\sqrt {{\Gamma _{B1}}} \left( {{R_{11}}\sqrt {{\Gamma _{C1}}} {e^{j{k_0}{l_{12}}}} + {R_{41}}\sqrt {{\Gamma _{C4}}} {e^{j{k_0}\left( {{l_{14}} + {l_{12}}} \right)}}} \right) - j\sqrt {{\Gamma _{B2}}} \left( {{R_{12}}\sqrt {{\Gamma _{C1}}}  + {R_{42}}\sqrt {{\Gamma _{C4}}} {e^{j{k_0}{l_{14}}}}} \right)
\end{dmath}

For initially photon propagating in the $C$ waveguide, transmission amplitudes are the following:
\begin{dmath}
{t_{CA}} =  - j\sqrt {{\Gamma _{C1}}} \left( \begin{array}{l}
{R_{21}}\sqrt {{\Gamma _{A2}}} {e^{j{k_0}{l_{12}}}}\\
 + {R_{31}}\sqrt {{\Gamma _{A3}}} {e^{j{k_0}\left( {{l_{12}} + {l_{23}}} \right)}}
\end{array} \right) - j\sqrt {{\Gamma _{C4}}} \left( \begin{array}{l}
{R_{24}}\sqrt {{\Gamma _{A2}}} {e^{j{k_0}\left( {{l_{14}} + {l_{12}}} \right)}}\\
 + {R_{34}}\sqrt {{\Gamma _{A3}}} {e^{j{k_0}\left( {{l_{14}} + {l_{12}} + {l_{23}}} \right)}}
\end{array} \right);
\end{dmath}
\begin{dmath}
{t_{CB}} =  - j\sqrt {{\Gamma _{C1}}} \left( {{R_{11}}\sqrt {{\Gamma _{B1}}}  + {R_{21}}\sqrt {{\Gamma _{B2}}} {e^{j{k_0}{l_{12}}}}} \right) - j\sqrt {{\Gamma _{C4}}} \left( {{R_{14}}\sqrt {{\Gamma _{B1}}} {e^{j{k_0}{l_{14}}}} + {R_{24}}\sqrt {{\Gamma _{B2}}} {e^{j{k_0}\left( {{l_{14}} + {l_{12}}} \right)}}} \right)
\end{dmath}
\begin{dmath}
{t_{CC}} = 1 - j\sqrt {{\Gamma _{C1}}} \left( {{R_{11}}\sqrt {{\Gamma _{C1}}}  + {R_{41}}\sqrt {{\Gamma _{C4}}} {e^{ - j{k_0}{l_{14}}}}} \right) - j\sqrt {{\Gamma _{C4}}} \left( {{R_{14}}\sqrt {{\Gamma _{C1}}} {e^{j{k_0}{l_{14}}}} + {R_{44}}\sqrt {{\Gamma _{C4}}} } \right)
\end{dmath}
and reflection amplitues:
\begin{dmath}
{r_{CA}} =  - j\sqrt {{\Gamma _{C1}}} \left( {{R_{21}}\sqrt {{\Gamma _{A2}}}  + {R_{31}}\sqrt {{\Gamma _{A3}}} {e^{ - j{k_0}{l_{23}}}}} \right) - j\sqrt {{\Gamma _{C4}}} \left( {{R_{24}}\sqrt {{\Gamma _{A2}}} {e^{j{k_0}{l_{14}}}} + {R_{34}}\sqrt {{\Gamma _{A3}}} {e^{j{k_0}\left( {{l_{14}} - {l_{23}}} \right)}}} \right)
\end{dmath}
\begin{dmath}
{r_{CB}} =  - j\sqrt {{\Gamma _{C1}}} \left( {{R_{11}}\sqrt {{\Gamma _{B1}}} {e^{j{k_0}{l_{12}}}} + {R_{21}}\sqrt {{\Gamma _{B2}}} } \right) - j\sqrt {{\Gamma _{C4}}} \left( {{R_{14}}\sqrt {{\Gamma _{B1}}} {e^{j{k_0}\left( {{l_{14}} + {l_{12}}} \right)}} + {R_{24}}\sqrt {{\Gamma _{B2}}} {e^{j{k_0}{l_{14}}}}} \right)
\end{dmath}
\begin{dmath}
{r_{CC}} =  - j\sqrt {{\Gamma _{C1}}} \left( \begin{array}{l}
{R_{11}}\sqrt {{\Gamma _{C1}}} {e^{j{k_0}{l_{12}}}}\\
 + {R_{41}}\sqrt {{\Gamma _{C4}}} {e^{j{k_0}\left( {{l_{14}} + {l_{12}}} \right)}}
\end{array} \right) - j\sqrt {{\Gamma _{C4}}} \left( \begin{array}{l}
{R_{14}}\sqrt {{\Gamma _{C1}}} {e^{j{k_0}\left( {{l_{14}} + {l_{12}}} \right)}}\\
 + {R_{44}}\sqrt {{\Gamma _{C4}}} {e^{j{k_0}\left( {2{l_{14}} + {l_{12}}} \right)}}
\end{array} \right)
\end{dmath}

\section{}
Here we present elements of inverse matrix $R$. Firstly, let’s rewrite the effective Hamiltonian in more simpler form:
\begin{dmath}
{\hat H_{eff}}\left( \omega  \right) = \left( {\begin{array}{*{20}{c}}
{{H_{11}}}&{{H_{12}}\left( \omega  \right)}&0&{{H_{14}}\left( \omega  \right)}\\
{{H_{12}}\left( \omega  \right)}&{{H_{22}}}&{{H_{23}}\left( \omega  \right)}&0\\
0&{{H_{23}}\left( \omega  \right)}&{{H_{33}}}&0\\
{{H_{14}}\left( \omega  \right)}&0&0&{{H_{44}}}
\end{array}} \right)
\end{dmath}
In these formulation of the effective hamiltionian inverse matrix $R$ elements are:
\begin{dmath}
{R_{11}}\left( \omega  \right) = \frac{1}{{D\left( \omega  \right)}}\left( {\omega  - {H_{44}}} \right)\left( {\left( {\omega  - {H_{33}}} \right)\left( {\omega  - {H_{22}}} \right) - H_{23}^2\left( \omega  \right)} \right)\\
\end{dmath}
\begin{dmath}
{R_{12}}\left( \omega  \right) = {R_{21}}\left( \omega  \right) = \frac{1}{{D\left( \omega  \right)}}{H_{12}}\left( \omega  \right)\left( {\omega  - {H_{33}}} \right)\left( {\omega  - {H_{44}}} \right)\\
\end{dmath}
\begin{dmath}
{R_{13}}\left( \omega  \right) = {R_{31}}\left( \omega  \right) = \frac{1}{{D\left( \omega  \right)}}{H_{12}}\left( \omega  \right){H_{23}}\left( \omega  \right)\left( {\omega  - {H_{44}}} \right)\\
\end{dmath}
\begin{dmath}
{R_{14}}\left( \omega  \right) = {R_{41}}\left( \omega  \right) = \frac{1}{{D\left( \omega  \right)}}{H_{14}}\left( \omega  \right)\left( {\left( {\omega  - {H_{33}}} \right)\left( {\omega  - {H_{22}}} \right) - H_{23}^2\left( \omega  \right)} \right)
 \end{dmath}
\begin{dmath}
{R_{22}}\left( \omega  \right) = \frac{1}{{D\left( \omega  \right)}}\left( {\omega  - {H_{33}}} \right)\left( {\left( {\omega  - {H_{11}}} \right)\left( {\omega  - {H_{44}}} \right) - H_{14}^2\left( \omega  \right)} \right)
 \end{dmath}
\begin{dmath}
{R_{23}}\left( \omega  \right) = {R_{32}}\left( \omega  \right) = \frac{1}{{D\left( \omega  \right)}}{H_{23}}\left( \omega  \right)\left( {\left( {\omega  - {H_{11}}} \right)\left( {\omega  - {H_{44}}} \right) - H_{14}^2\left( \omega  \right)} \right)
 \end{dmath}
\begin{dmath}
{R_{24}}\left( \omega  \right) = {R_{42}}\left( \omega  \right) = \frac{1}{{D\left( \omega  \right)}}{H_{12}}\left( \omega  \right){H_{14}}\left( \omega  \right)\left( {\omega  - {H_{33}}} \right)
 \end{dmath}
\begin{dmath}
{R_{33}}\left( \omega  \right) = \frac{1}{{D\left( \omega  \right)}}\left[ \begin{array}{l}
\left( {\omega  - {H_{11}}} \right)\left( {\omega  - {H_{22}}} \right)\left( {\omega  - {H_{44}}} \right)\\
 - \left( {\omega  - {H_{22}}} \right)H_{14}^2\left( \omega  \right) - \left( {\omega  - {H_{44}}} \right)H_{12}^2\left( \omega  \right)
\end{array} \right]
 \end{dmath}
\begin{dmath}
{R_{34}}\left( \omega  \right) = {R_{43}}\left( \omega  \right) = \frac{1}{{D\left( \omega  \right)}}{H_{12}}\left( \omega  \right){H_{14}}\left( \omega  \right){H_{23}}\left( \omega  \right)
 \end{dmath}
\begin{dmath}
{R_{44}}\left( \omega  \right) = \frac{1}{{D\left( \omega  \right)}}\left[ \begin{array}{l}
\left( {\omega  - {H_{11}}} \right)\left( {\omega  - {H_{22}}} \right)\left( {\omega  - {H_{33}}} \right)\\
 - \left( {\omega  - {H_{11}}} \right)H_{23}^2\left( \omega  \right) - \left( {\omega  - {H_{33}}} \right)H_{12}^2\left( \omega  \right)
\end{array} \right]
 \end{dmath}
where  is determinant of R matrix, and can be written as follows:
 \begin{dmath*}
\begin{array}{l}
D\left( \omega  \right) = H_{14}^2\left( \omega  \right)H_{23}^2\left( \omega  \right)\\
 - \left( {\omega  - {H_{44}}} \right)\left[ {\left( {\omega  - {H_{11}}} \right)H_{23}^2\left( \omega  \right) + \left( {\omega  - {H_{33}}} \right)H_{12}^2\left( \omega  \right)} \right]\\
 + \left( {\omega  - {H_{22}}} \right)\left( {\omega  - {H_{33}}} \right)\left[ {\left( {\omega  - {H_{11}}} \right)\left( {\omega  - {H_{44}}} \right) - H_{14}^2\left( \omega  \right)} \right]
\end{array}
\end{dmath*}

\end{document}